\documentclass[conference]{IEEEtran}
\IEEEoverridecommandlockouts

\usepackage{cite}
\usepackage{amsmath,amssymb,amsfonts}
\usepackage{algorithmic}
\usepackage{graphicx}
\usepackage{tabularx}
\usepackage{textcomp}
\usepackage{xcolor}
\usepackage{lscape} 
\usepackage{booktabs}
\usepackage{makecell}
\usepackage{threeparttable}
\usepackage{array,multirow,graphicx}
\usepackage{float}
\usepackage{multicol}
\usepackage{etoolbox}
\usepackage{float}
\usepackage{hyperref}

\usepackage{adjustbox}
\usepackage{array}

\newcolumntype{R}[2]{%
    >{\adjustbox{angle=#1,lap=\width-(#2)}\bgroup}%
    l%
    <{\egroup}%
}

\def\BibTeX{{\rm B\kern-.05em{\sc i\kern-.025em b}\kern-.08em
    T\kern-.1667em\lower.7ex\hbox{E}\kern-.125emX}}

\makeatletter
\patchcmd{\@maketitle}
  {\addvspace{6pt}}
  {\begin{multicols}{4}\centering}
  {}{}
\patchcmd{\@maketitle}
  {\par\nobreak \vskip 12pt}
  {\end{multicols}\par\nobreak \vskip 12pt}
  {}{}
\makeatother

\makeatletter
\newcommand{\linebreakand}{%
  \end{@IEEEauthorhalign}
  \hfill\mbox{}\par
  \mbox{}\hfill\begin{@IEEEauthorhalign}
}
\makeatother

\renewcommand{\footnoterule}{%
\kern -3pt
\hrule width 2in
\kern 2.6pt
}
\begin{document}

\title{Future You: A Conversation with an AI-Generated Future Self Reduces Anxiety, Negative Emotions, and Increases Future Self-Continuity}









\DeclareRobustCommand*{\IEEEauthorrefmark}[1]{%
  \raisebox{0pt}[0pt][0pt]{\textsuperscript{\footnotesize #1}}%
}

\author{
\IEEEauthorblockN{
Pat Pataranutaporn*\IEEEauthorrefmark{1},
Kavin Winson*\IEEEauthorrefmark{2},
Peggy Yin*\IEEEauthorrefmark{3},
Auttasak Lapapirojn\IEEEauthorrefmark{2},\\
Pichayoot Ouppaphan\IEEEauthorrefmark{2},
Monchai Lertsutthiwong\IEEEauthorrefmark{2},
Pattie Maes\IEEEauthorrefmark{1},
Hal E. Hershfield\IEEEauthorrefmark{4}
}

\IEEEauthorblockA{\IEEEauthorrefmark{1}MIT Media Lab,
Massachusetts Institute of Technology,
Cambridge, MA, United States\\
Email: \{patpat, pattie\}@media.mit.edu}

\IEEEauthorblockA{\IEEEauthorrefmark{2}KASIKORN Labs,
KASIKORN Business-Technology Group,
Nonthaburi, Thailand\\
Email: \{kavin.w, auttasak.l, pichayoot.o, monchai.le\}@kbtg.tech}

\IEEEauthorblockA{\IEEEauthorrefmark{3}Harvard University,
Cambridge, MA, United States\\
Email: pyin@college.harvard.edu}

\IEEEauthorblockA{\IEEEauthorrefmark{4}Anderson School of Management,
University of California, Los Angeles,
CA, United States\\
Email: hal.hershfield@anderson.ucla.edu}
}

\maketitle
\def\thefootnote{*}\footnotetext{These authors contributed equally to this work.}\def\thefootnote{\arabic{footnote}}

\begin{figure}[h]
\sbox0{
\begin{minipage}{\textwidth}
    \centering
    \includegraphics[width=1.0\linewidth]{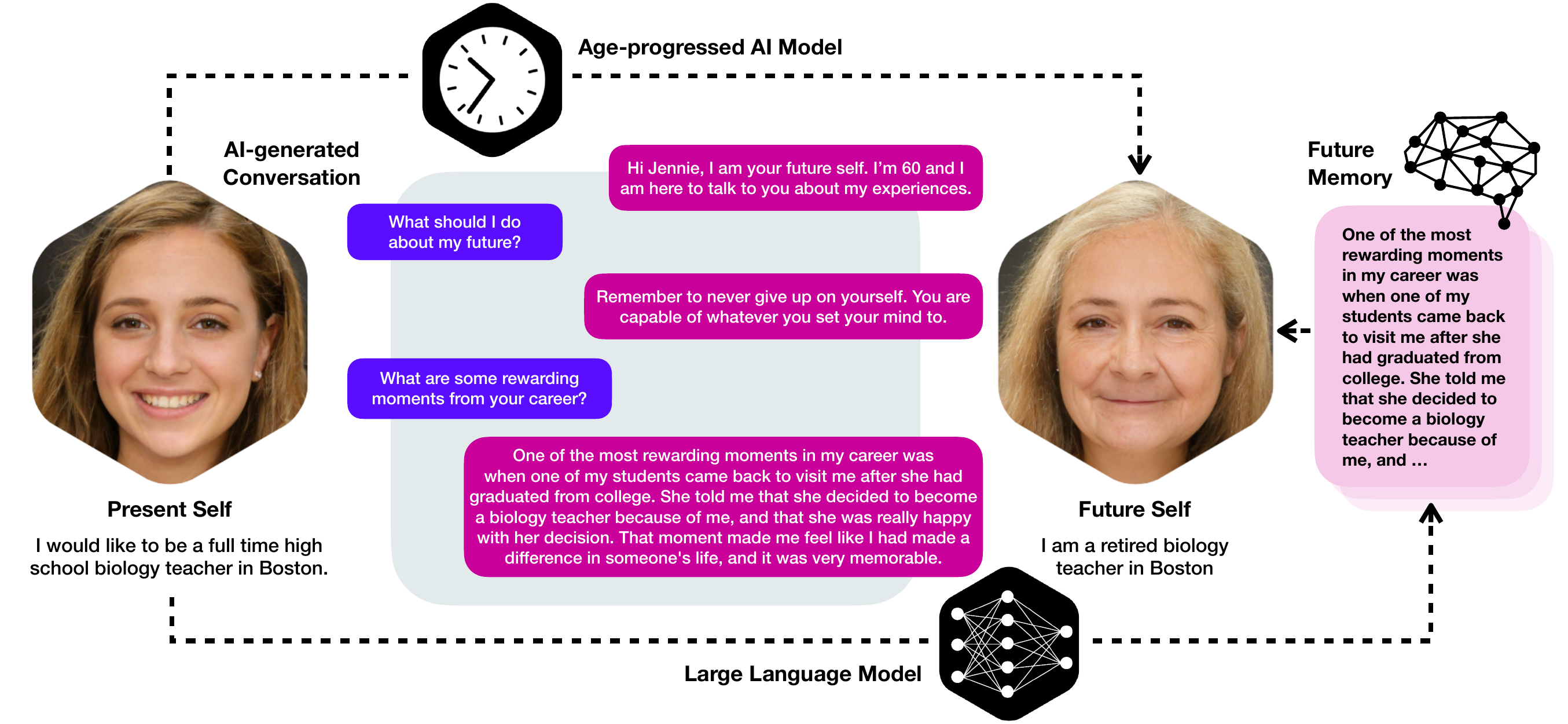}
    \caption{“Future You” is an interactive chat platform that allows users to converse with a relatable yet virtual version of their future selves in real-time. This is achieved through a large language model that has been personalized based on a pre-intervention survey focused on users' future goals and personal qualities. To make the conversation realistic, the system generates an individualized “future memory” for each user's future self, containing a backstory for the user at age 60. To increase the believability of the future self character, the system age-progresses the user's submitted headshot.}
    \label{fig:teaser figure}
\end{minipage}}%
\rlap{\usebox0}
\end{figure}

\begin{abstract}
We introduce “Future You,” an interactive, brief, single-session, digital chat intervention designed to improve future self-continuity—the degree of connection an individual feels with a temporally distant future self—a characteristic that is positively related to mental health and wellbeing. Our system allows users to chat with a relatable yet AI-powered virtual version of their future selves that is tuned to their future goals and personal qualities. To make the conversation realistic, the system generates a “future memory”—a unique backstory for each user—to \vspace{10.08cm}\hspace{11cm}
\\ create a throughline between the user’s present age (between 18-30) and their life at age 60. The “Future You” character also adopts the persona of an age-progressed image of the user. In our preregistered study (N = 344), we found that after a brief interaction with the “Future You” character, users reported significantly decreased anxiety and increased future self-continuity compared to control conditions. This is the first study successfully demonstrating the use of personalized AI-generated characters to improve users’ future self-continuity and wellbeing.
\end{abstract}


\section{Introduction}

\noindent\textit{Imagine yourself in 30 years...}\newline
\textit{How vividly can you imagine your future self?}\newline
\textit{How similar does your future self feel to your present self?}\newline
\textit{How positively do you regard your future self?}\newline


Those who can imagine their future selves vividly, regard their future selves positively, or consider their future selves to be similar to their present self exhibit a strong sense of future self-continuity, the degree of connectedness one feels with a temporally distant future self \cite{ersner2009don}. An emerging field of study on future self-continuity has shown that high future self-continuity promotes better saving behavior, academic performance, mental health, and subjective quality of life \cite{ersner2009don, hershfield2011future, van2013vividness, sokol2019development}. Current future self-continuity interventions help participants embody the perspective of their future selves via methods like letter writing, method acting, and virtual reality (VR) simulations \cite{ganschow2021looking, pronin2006temporal, chishima2021conversation}. These interventions have shown that embodying a temporally-distanced perspective—e.g., focusing on the distant future—can not only improve future self-continuity but also reduce anxious and overwhelming feelings \cite{chishima2021temporal}. However, these interventions require individuals to deeply reflect upon potentially ambiguous core identities or actively imagine a version of themselves that they have not yet experienced—tasks that can be particularly challenging given that humans are naturally biased toward the preferences of our present selves \cite{quoidbach2013end}. Moreover, these embodiment-based interventions require individuals to have access to a VR headset, undergo training in the method acting technique, or perform physical activities in order to effectively switch into the perspective of their future self, making them inaccessible to many people.

We introduce “Future You,” an interactive, digital chat intervention designed to improve future self-continuity and wellbeing. In contrast, our approach uses AI-generated characters (realistic digital representations of a person)\cite{pataranutaporn2021ai, pataranutaporn2022ai, prasongpongchai2024interactive, danry2022ai} to facilitate exploration of future selves via interaction rather than embodiment. In intimate conversational settings, AI-generated characters that act in a human-like way have been able to encourage individuals to both reveal more authentic information about themselves than they would when talking with actual humans, and undertake positive behavioral and cognitive changes \cite{lucas2014s}. These advancements in visual technologies have been leveraged for a variety of use cases, including generating facial animations \cite {hong2022depth, yin2022styleheat, zakharov2019few, yao2022dfa, liu2022semantic, guo2021ad},  protecting people's privacy in documentaries and interviews \cite{chechnya}, dubbing of films \cite{takahashi_2020}, and reanimating historical images \cite{dali, steinhauer2022nostalgia, living_memory}. Our research builds upon this recent progress in AI to generate highly humanlike characters impersonating a highly realistic visual representation of a user in the future.
 
Additionally, many chatbots have already been developed for providing effective mental health interventions. For example, one of the most widely adopted mental health chatbots is Woebot, an automated conversational agent that uses cognitive behavior therapy to help users monitor their mood \cite{nyt}. In a study with 70 individuals between 18 and 28, Woebot significantly reduced symptoms of depression, compared to a control group \cite{fitzpatrick2017delivering}. The study also found that participants in the Woebot group engaged with the conversational agent on daily or almost daily basis. However, while conversational agents offer feasible, engaging, and effective ways to deliver mental health support, democratizing access for those who have traditionally been reluctant to seek in-person mental health advice due to stigmatization or inaccessibility, the majority of these bots are rule-based rather than AI-generated and thus can only focus on specific conditions such as depression and autism \cite{abd2019overview,lucas2017reporting, vaidyam2019chatbots,fitzpatrick2017delivering}. In our research, we apply advances in AI-generated conversations based on large language models (LLM) and natural language processing (NLP) that make it possible to simulate natural-sounding conversations with a future self, aiming to investigate the use of AI-powered virtual conversational agents to improve mental health and wellbeing by increasing a user's sense of future self-continuity. We focused on the following research questions:

\begin{itemize}
\item How does having a conversation with an AI-generated future self affect negative and positive affect?
\item How does having a conversation with an AI-generated future self affect future self-continuity?
\item How does having a conversation with an AI-generated future self affect other future-oriented and self-oriented outcome variables such as optimism for the future, agency towards goals, consideration for future consequences, self-reflection, and self-esteem? 
\end{itemize}

To our knowledge, this is the first work to demonstrate the use of AI-generated characters as an accessible and effective, interactive future self intervention. Here we summarize our contributions:

\begin{itemize}
  \item We introduce an accessible and interactive web-based intervention that uses AI to simulate a conversation with the future self.
  \item We present a method of using AI to generate future memories in order to create believable narratives told from the perspective of a user's future self.
  \item We demonstrate that our AI-based intervention can lower anxiety and boost future self-continuity.
\end{itemize}

\section{Related Work}
Our research is situated in the area of AI-generated characters, conversational agents, and interventions for increasing future self-continuity. So far, researchers have characterized two types of interventions, reflective and presentational, that can help strengthen future self-continuity. 

\subsubsection{Reflective Intervention}
Using the reflective intervention, participants are asked to spend time thinking about their future. In one study, participants were randomly assigned to send a letter to their future self, send a letter to their present self from the perspective of the future self, or write a letter about their daily life (the control condition) \cite{chishima2021temporal}. Participants in both letter-writing conditions showed an immediate decrease in negative affect and an increase in positive affect relative to the control condition. In another related study, researchers demonstrated that students who wrote a letter to their future self and responded to that letter as their future self showed increases in future self-continuity, career planning, and academically-related delay of gratification compared to students who only wrote a letter to their future selves \cite{chishima2021conversation}. These studies demonstrate how future perspective-taking and future self-continuity is not only related to mental wellbeing, but also career and academic outcomes. However, even though reflective interventions can be easily be deployed at scale with low technological requirements, they are limited by their heavy reliance on the individual's ability to imagine, which requires mental effort and may vary significantly from person to person.

\subsubsection{Presentational Intervention}
A presentational intervention provides a visual representation of the participant's future self, typically in a form of an age-processed image or avatar, producing a similar behavioral change in the participant without requiring active reflection. For example, participants viewing realistic computer renderings of their future self using immersive virtual reality hardware and interactive decision aids exhibited an increased tendency to accept later monetary rewards over immediate ones \cite{hershfield2011increasing}. A related study exposing college students to age-progressed avatars multiple times a day resulted in students allocating more money for saving and receiving a higher score on a subsequent financial quiz than participants who did not view age-progressed images of themselves \cite{sims2020future}. Similarly, when banking customers were exposed to images of their future selves, they were 16\% more likely to make a contribution to a retirement account compared to those in a control condition \cite{hal_rec}. Recently, a smartphone-based intervention was developed to enhance future-oriented thinking and behavior by strengthening future self-identification. Participants take a ‘selfie’ during the intake using the integrated camera of the participant's smartphone. This image is then digitally age-progressed by approximately 10 years using a custom-made server in conjunction with the online service Change My Face. The aged image is then converted into a 3D digital representation, creating an avatar of the subject's projected future self \cite{mertens2022future}. A randomized controlled pilot study with first-year university students evaluated its immediate and long-term efficacy, as well as effects after each module. Results showed an initial decrease in goal commitment but trends toward increased future orientation and self-efficacy at a 3-month follow-up. Additionally, the first module positively affected future self vividness. Such findings suggest the mobile potential to boost future-oriented cognition and behavior \cite{mertens2024novel}. Especially vivid renderings of future selves can lead to improved long-term decision-making.

Beyond direct exposure to a future self visualization, researchers have also explored the use of age-progressed avatars for roleplaying exercises. In one study \cite{van2022interaction}, convicted offenders interacted with an age-progressed avatar representing their future self in virtual reality, reflecting on their current lifestyle while alternating between the perspective of their present self and their future self. The exercise increased the vividness of participants' future selves to them compared to baseline, resulting in reduced alcohol use and overspending one week later. Another study used virtual reality to facilitate participants roleplaying as their successful future selves; participants answered questions about what it felt like to become their successful future self and the path they took to get there\cite{ganschow2021looking}. This exercise was conducted in a virtual reality environment to investigate the possible added value of the virtual environment with respect to improved focus, perspective-taking, and effectiveness, especially for participants with lower imagination. Results indicated that the perspective-taking exercise in virtual reality substantially increased all domains of future self-continuity (similarity, vividness, and liking), while the in vivo equivalent increased only liking and vividness. These findings show that the perspective taking exercise in a VR environment can reliably increase the future self-continuity domains. However, such interventions are limited by accessibility, as virtual reality headsets remain inaccessible to the majority of people.

\begin{figure*}
    \centering
    \includegraphics[width=1\linewidth]{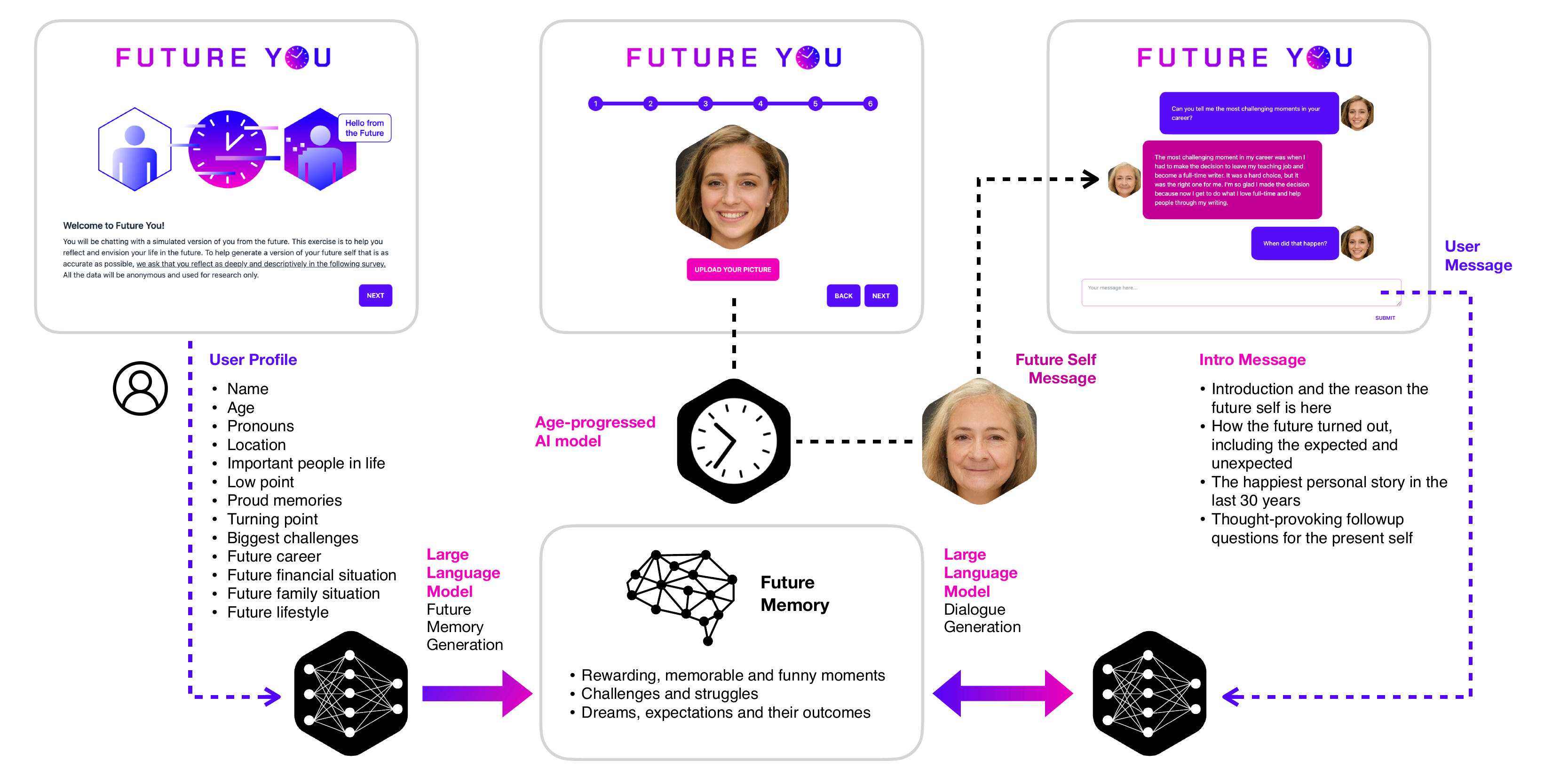}
    \caption{Our Future You system consists of four modules: Life Story Interface, Age-Progressed AI, Future Memory Architecture, and Chat Interface.}
    \label{fig:futureyou-system-arch}
\end{figure*}

\section{Methodology}
Here, we present a third approach that is scalable, web-based, and allows users to interactively generate a version of their future selves without the demand for a vivid imagination or embodied exercises. Our research develops an accessible and effective future self-continuity intervention that allows users to interact with a realistic and relatable future self as a conversational partner via a large language model (GPT-3.5). We used linguistic cues to signal continuity and similarity between the user's future and present self (e.g. saying “when I was your age...” and noting similarities between their perception of their past goals and the user's present goals). We also presented users with a vivid, narrated, and visualized version of their future selves via the sharing of future memories and an age-progressed photo. In doing so, we hoped to increase the accessibility of this future self intervention by eliminating the need for the user to construct their continuous future self image and story by themselves. Our Future You system consists of four modules, as shown in Figure \ref{fig:futureyou-system-arch}:

\subsubsection{Life Story Interface}
First, the user is prompted to answer a series of demographic, life-narrative, and goals-oriented questions about themselves. This information is used to generate an accurate future self simulation of the person. The survey is available to participants as individual questions in a sequential format embedded in a web interface built with JavaScript. Each question has its own free text input box for the user to answer, with a default grey example response to the question. The questions are separated into two main categories. The first set of questions focuses on the user's present, asking for basic details such as the user's name, age, pronouns, location, and the people in their lives who are most important to them. Drawing from the Life Story Interview, we also ask the user to share past experiences that serve as turning points, proud points, and low points for them \cite{mcadams2007life}. After this, they enter the second phase of the questionnaire, which prompts them to answer questions regarding their career and professional accomplishments, financial status, family life, and personal life outcomes they would like for themselves at age 60. Participants are instructed to reflect deeply and descriptively as they answer questions about themselves in the survey.

\subsubsection{Age-progressed AI}
After the user inputs their information, they are directed to an interface to upload their portrait from local system storage. We use an AI-powered image editing model called StyleCLIP \cite{patashnik2021styleclip} to age the portrait, simulating how the user might look as a 60-year-old version of themselves. This process adds features like wrinkles and white hair to create an older appearance. The aging process takes only a few seconds, during which a time-warping animation plays. Upon success, the enlarged final image is shown to the user, suggesting it is what their future self might look like. The user can take as long as they want to observe their 60-year-old self before proceeding to the next step. 

\subsubsection{Future Memory Architecture}
In order to generate an accurate and realistic future self, the language model uses the input data from the survey to generate a backstory of the user's personal history from the perspective of their 60-year-old self. This future memory provides the future self a continuous past and present experience to draw from, ensuring that the generated responses from the future self present as a cohesive narrative and feel personalized to each user. The initial future memory of the future self character is generated by incorporating the user's survey inputs into the following template prompt:

\begin{quotation}
The following is the interview of \textit{name}, who is a successful \textit{career}. \textit{name}'s pronoun and sexual orientation are \textit{pronoun\_and\_sexual\_orientation}. \textit{name} is from \textit{place}. The most important people in \textit{name}'s life are: “\textit{people\_in\_life}”. Right now, \textit{name} is 60 years old and can share insightful stories and experiences, give definitive advice and life lessons as \textit{name} reflects on life. In the past, the most important low point in \textit{name}'s life was “\textit{low\_point}”. \textit{name} also experienced a turning point in their life when “\textit{turning\_point}”. \textit{name} has dedicated their life to a significant life project called “\textit{life\_project}”. \textit{name} is also proud of great things that the young \textit{name} has done: “\textit{proud}”. In the past, when \textit{name} was \textit{age} years old, \textit{name} had many dreams and hopes for the future. \textit{age}-year-old \textit{name} has said “\textit{professional\_accomplish}, \textit{financial\_accomplish}, and \textit{family\_accomplish}”. Right now, \textit{name} is living in \textit{where\_to\_live} and having the following daily life: \textit{daily\_life}.
\end{quotation}

Then, for each input topic, we append probing questions to the user's statements as prompts to GPT-3.5 to generate a richer set of individual memories. Each individual memory is generated in parallel to minimize the processing time when loading the chat interface. Upon completion, all memory fragments are combined to form a robust backstory for the virtual character. For example, a user might answer in a survey regarding a future career that: \begin{quote} “I would like to be a full-time high school biology teacher in Boston. I am very excited to teach kids and help them learn new things about the natural world.”
\end{quote}

The future memory module would then use the information to generate a memory for the user as a retired biology teacher at age 60:
\begin{quote}
“A rewarding story from my career would be the time when I was able to help a struggling student turn their grades around and pass their biology class. It was so gratifying to see the student's face light up with pride and accomplishment. Another memorable moment was taking my students on a field trip to a local nature preserve. They were so excited to explore and learn about all the different plants and animals.
”\end{quote}

\subsubsection{Chat Interface}
Before entering the chat interface, participants are informed that they will be chatting with a simulated version of their future selves. Participants are then brought to the chat interface powered by a JavaScript API to chat with their AI-generated 60-year-old future self. The system is designed with the user interface of a traditional messaging application with a scrollable area for the chronological history of text bubbles and an input message box to type and send messages at the bottom of the page. Within the chat interface, the future self has the age-progressed image as a profile picture, while the user is represented by the original image they uploaded during the age-progression module. The future self is then prompted with a series of questions to generate a starting series of messages demonstrating its functionality and future memories:

\begin{itemize}
    \item Can you casually introduce yourself, your name, and your age, which is 60 years old, and why you are here? Casually and briefly mention that the future might be different than you expect and mention that your future might be different.
    \item Please briefly tell me what your dream was with “when I was your age...” and how it turned out to be. What are the things that you expect and didn't expect?
    \item Please tell me the happiest stories about your family as you reflected in the last 30 years, starting with “You know, when I think of my life...”, and share insightful motivation for my future.
    \item Reflecting on past experiences, what and how has the life project you have been involved in deeply impacted you and others in a genuine and heartfelt way? How did you initially become involved in this project, and how has it developed over time? Furthermore, why do you believe this project holds such importance for both yourself and the individuals it has touched?
\end{itemize}

The participants are then allowed to chat with their future selves at will by typing in the message box and pressing “Enter” or clicking the “Send” button. The message is displayed, and the future self’s response is generated accordingly via an underlying API call. As the AI is outputting a response, a typing animation is shown to simulate standard texting interfaces. After 16 exchanged messages, a non-intrusive clickable button appears at the bottom of the application for the user to thank their future self and move on if they are finished. The user can also ignore the button to continue with the chat. The messages exchanged with the AI are sent to a Google Sheet for further analysis.

\section{Experiment}
We conducted a study with 400 participants using Qualtrics, an online survey platform. After excluding participants who failed attention checks or experienced technical issues, 344 participants remained in the final sample (51.16\% male, 44.19\% female, 3.78\% nonbinary, 0.58\% self-described, 0.29\% preferred not to say; $age_M$ = 25.05, $age_SD$ = 3.46). We distributed the study on CloudResearch. The study was set to be balanced between male and female participants aged 18-30 years old, who were pre-screened to be fluent in English. The participants were asked to consent to have their conversation and survey data used anonymously prior to proceeding to the rest of the survey. Participants completed a pre-intervention questionnaire based on established psychological scales. They were then randomly assigned to one of three control conditions or an experimental condition:

\begin{enumerate}
    \item \textbf{Future You}: An experimental condition in which participants fill out the autobiographical survey and chat with an AI-generated future self for a minimum of 10 minutes and a maximum of 30 minutes
    \item \textbf{Control}: A neutral control condition in which participants only fill out pre- and post-intervention surveys
    \item \textbf{Chat}: An active control condition in which participants interact with a generic virtual assistant chatbot
    \item \textbf{Questionnaire}: An active control condition in which participants complete the autobiographical survey but do not interact with the AI-generated future self
\end{enumerate}

After the conversation, all participants were asked to complete a post-intervention questionnaire about their experience. This survey included psychological scales, questions about any technical issues encountered, attention checks, and an open-ended response allowing participants to provide additional feedback or comments. We also collected demographic information, including gender, sexual orientation, age, education level, race, and ethnicity. 

\subsection{Measurements}
This study adapted items from eight established psychological scales: an abridged Emotion and Arousal Checklist \cite{oda2015development}, State Optimism Measure \cite{millstein2019development}, the Self Reflection and Insight Scale \cite{silvia2022self}, and Future Self-Continuity Questionnaire \cite{sokol2019development}, adapted items from the Adult Hope Scale \cite{snyder2007adult}, Consideration for Future Consequences scale \cite{hevey2010consideration}, Rosenberg's Self-Esteem Scale \cite{martin2007rosenberg}, and a custom perceived realism questionnaire. All scales were presented as 7-point Likert scales for consistency. For the future self-continuity Questionnaire, we adapted the similarity questions to reflect on the user's perspective from their 60-year-old self (e.g. “How similar are you now to what you will be like when you are 60 years old?”), and for the likability and vividness questions, we adapted the questions to reflect a 10-year-projection (e.g. “How vividly can you imagine what you will be like in 10 years from now?”). Participants completed these measures both before and after the intervention to assess any changes in psychological constructs resulting from the intervention.

\subsection{Analysis}
We calculated the differences between pre-intervention and post-intervention questionnaire results for each construct. For the emotions measure, we analyzed emotions individually, as well as overall positive and negative composites. We then compared the composite scores and individual scores between conditions. For the analysis, we first assessed if the normality assumption was met for each outcome variable distribution using the Shapiro-Wilk test. If the normality assumption was not met, we performed a Kruskal-Wallis test followed by a post-hoc Dunn test using the Bonferroni error correction. If the normality assumption was met, we then conducted a homogeneity test using a Levene test to assess whether the samples were from populations with equal variances. If the samples were not homogeneous, we ran a Welch analysis of variance (ANOVA) and a Tukey's honestly significant difference test (Tukey's HSD) test. If the samples were homogeneous, we ran a basic ANOVA test with a Tukey post-hoc test. 

\subsection{Approvals} 
This research was reviewed and approved by the MIT Committee on the Use of Humans as Experimental Subjects, protocol number E-4261. The study was also preregistered at AsPredicted (\#157535).

\section{Results}
\begin{figure*}[t!] 
\centering
\includegraphics[width=1\linewidth]{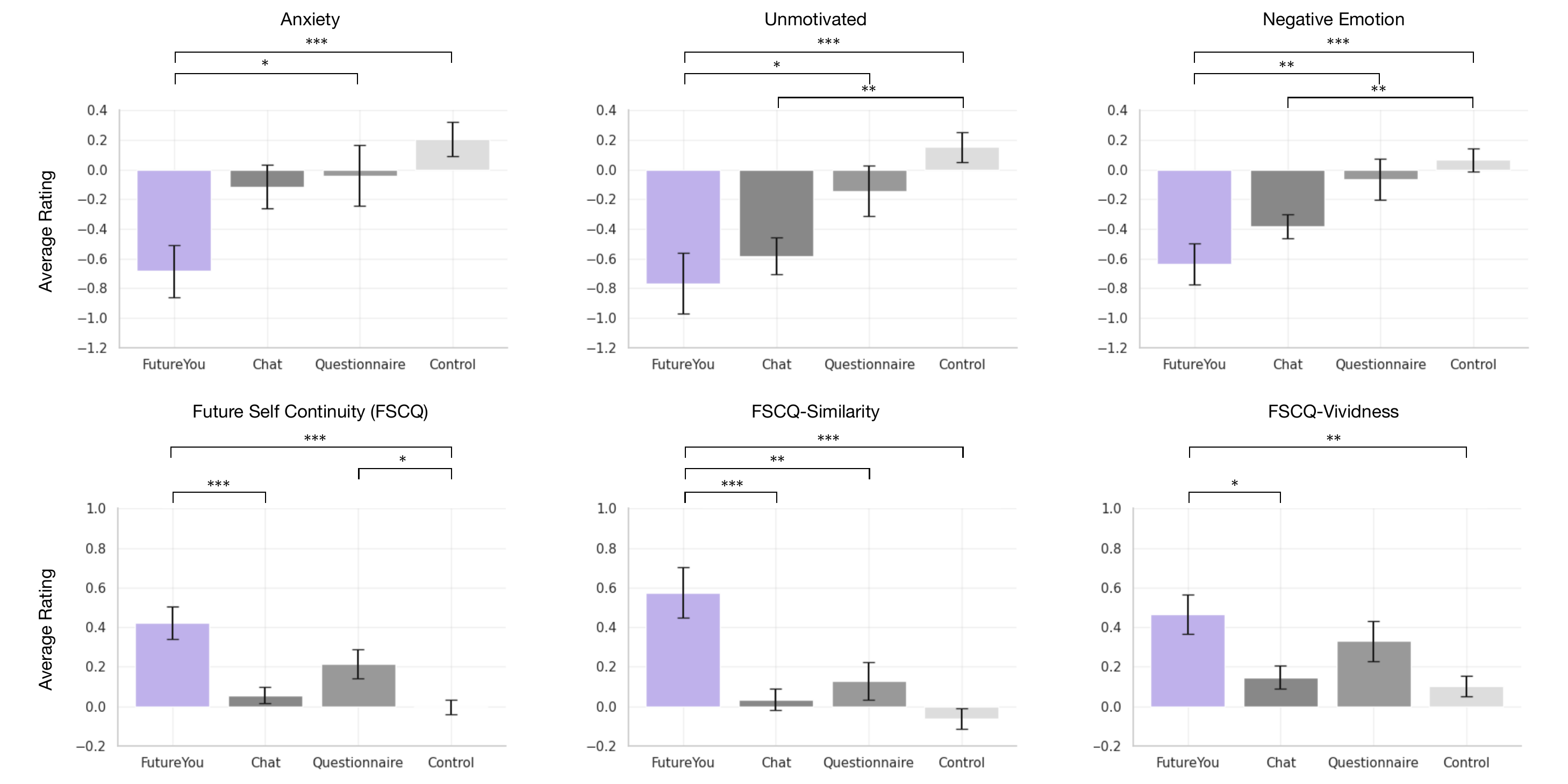}

\caption{After interacting with the “Future You” character, users reported reduced negative emotions such as anxiety and unmotivated feelings, and increased future self-continuity.}
 \label{fig:futureyou-graph}
\end{figure*}

The present study investigated the effects of different intervention conditions on various wellbeing and psychological outcomes, including emotions, future self-continuity, agency, optimism, future consideration, self-esteem, self-reflection, and insight. Participants were randomly assigned to one of four conditions: Future You (73 participants), Questionnaire (76 participants), Chat (103 participants), or Control (92 participants). After interacting with the Future You character, users reported reduced negative emotions such as anxiety and unmotivated feelings and increased future self-continuity, as shown in Figure \ref{fig:futureyou-graph} and Table \ref{tab:anova_no_posthoc}.

\begin{table*}[h]
\scriptsize
\begin{tabularx}{\textwidth}{Xcccccccc}
\textbf{Measure} & \textbf{Homogeneity} & \textbf{ANOVA Type} & \textbf{F-statistic} & \textbf{p-value} & \textbf{Future You} & \textbf{Chat} & \textbf{Questionnaire} & \textbf{Control} \\ \hline
\(\Delta\) Positive Emotion & No & Welch & 2.321 & 0.0769 & -0.20 \(\pm\) 0.99 & 0.01 \(\pm\) 0.67 & -0.18 \(\pm\) 0.79 & 0.05 \(\pm\) 0.52 \\
\(\Delta\) Negative Emotion & No & Welch & 8.835 & 1.76e-05$^{****}$ & -0.63 \(\pm\) 1.20 & -0.38 \(\pm\) 0.82 & -0.07 \(\pm\) 1.19 & 0.07 \(\pm\) 0.77 \\  
\(\Delta\) Anxious & Yes & One-way & 5.134 & 0.0017$^{**}$ & -0.68 \(\pm\) 1.52 & -0.12 \(\pm\) 1.50 & -0.04 \(\pm\) 1.77 & 0.21 \(\pm\) 1.10 \\
\(\Delta\) Overwhelmed & No & Welch & 2.245 & 0.0848 & -0.45 \(\pm\) 1.74 & -0.45 \(\pm\) 1.10 & -0.01 \(\pm\) 1.36 & -0.16 \(\pm\) 1.26 \\
\(\Delta\) Unmotivated & No & Welch & 9.618 & 6.58e-06$^{****}$ & -0.77 \(\pm\) 1.75 & -0.58 \(\pm\) 1.26 & -0.14 \(\pm\) 1.48 & 0.15 \(\pm\) 0.98 \\
\(\Delta\) Agency & No & Welch & 4.677 & 0.0036$^{**}$ & 0.10 \(\pm\) 0.79 & 0.20 \(\pm\) 0.67 & 0.00 \(\pm\) 0.83 & -0.12 \(\pm\) 0.57 \\
\(\Delta\) Optimism & No & Welch & 1.497 & 0.2171 & 0.24 \(\pm\) 0.89 & 0.26 \(\pm\) 0.54 & 0.14 \(\pm\) 0.90 & 0.09 \(\pm\) 0.66 \\
\(\Delta\) FSCQ 1 (Similarity) & No & Welch & 7.446 & 0.0001$^{***}$ & 0.58 \(\pm\) 1.08 & 0.03 \(\pm\) 0.56 & 0.13 \(\pm\) 0.83 & -0.06 \(\pm\) 0.52 \\ 
\(\Delta\) FSCQ 2 (Vividness) & No & Welch & 4.377 & 0.0054$^{**}$ & 0.47 \(\pm\) 0.84 & 0.15 \(\pm\) 0.59 & 0.33 \(\pm\) 0.90 & 0.10 \(\pm\) 0.49 \\
\(\Delta\) FSCQ 3 (Positivity) & No & Welch & 2.551 & 0.0573 & 0.22 \(\pm\) 0.88 & -0.01 \(\pm\) 0.68 & 0.19 \(\pm\) 0.92 & -0.05 \(\pm\) 0.60 \\ 
\(\Delta\) Future Self-Continuity & No & Welch & 8.571 & 2.49e-05$^{****}$ & 0.42 \(\pm\) 0.70 & 0.06 \(\pm\) 0.42 & 0.22 \(\pm\) 0.64 & -0.00 \(\pm\) 0.35 \\
\(\Delta\) Future Consideration & No & Welch & 1.192 & 0.3145 & -0.01 \(\pm\) 0.63 & 0.09 \(\pm\) 0.50 & 0.19 \(\pm\) 0.75 & 0.04 \(\pm\) 0.62 \\ 
\(\Delta\) Self-Esteem & No & Welch & 1.209 & 0.3079 & 0.10 \(\pm\) 0.76 & 0.10 \(\pm\) 0.55 & 0.06 \(\pm\) 0.75 & -0.03 \(\pm\) 0.45 \\
\(\Delta\) Self-Reflection & Yes & One-way & 0.471 & 0.7030 & -0.04 \(\pm\) 0.82 & 0.02 \(\pm\) 0.55 & 0.08 \(\pm\) 0.75 & -0.03 \(\pm\) 0.70 \\
\(\Delta\) Insight & Yes & One-way & 0.692 & 0.5574 & 0.03 \(\pm\) 0.73 & 0.04 \(\pm\) 0.68 & -0.07 \(\pm\) 0.66 & -0.07 \(\pm\) 0.65 \\
\multicolumn{9}{l}{$^{*}$p$<$0.05; $^{**}$p$<$0.01; $^{***}$p$<$0.001; $^{****}$p$<$0.0001} \\
\end{tabularx}

\caption{ANOVA results without post-hoc analysis. Measures violating the equal variances assumption used Welch's ANOVA. Means (M) are provided for each condition.}
\label{tab:anova_no_posthoc}
\end{table*}

\textbf{\(\Delta\) Negative Emotion:} A Welch one-way ANOVA was conducted due to violation of the homogeneity of variances assumption (Levene's test, p=0.004). There was a significant effect of intervention conditions on change in negative emotion at the p$<$.001 level [F(3,181.60)=8.84, p$<$0.001]. Post hoc comparisons using the Tukey HSD test indicated that the mean change for the Future You condition (M=-0.63, SD=1.20) was significantly different than the Control condition (M=0.07, SD=0.77, p=0.001) and the Questionnaire condition (M=-0.07, SD=1.19, p=0.003). The Chat condition (M=-0.38, SD=0.82) also differed significantly from Control (p=0.009).

\textbf{\(\Delta\) Anxious}: A one-way ANOVA was conducted to compare the effect of intervention conditions on changes in anxiety levels. The assumption of homogeneity of variances was met (Levene's test, p=0.068). There was a significant effect of intervention conditions on change in anxiety at the p$<$.01 level [F(3,376)=5.13, p=0.002]. Post hoc comparisons using the Tukey test indicated that the mean change in anxiety for the Future You condition (M=-0.68, SD=1.52) was significantly different than the Control condition (M=0.21, SD=1.10, p=0.001) and the Questionnaire condition (M=-0.04, SD=1.77, p=0.040), and marginally different than the Chat condition (M=-0.12, SD=1.50, p=0.059). 

\textbf{\(\Delta\) Unmotivated}: A Welch one-way ANOVA was conducted due to a violation of the homogeneity of variances assumption (Levene's test, p=0.002). There was a significant effect of intervention conditions on change in feeling unmotivated at the p$<$.001 level [F(3,182.92)=9.62, p$<$0.001]. Post hoc comparisons using the Tukey HSD test indicated that the mean change for the Future You condition (M=-0.77, SD=1.75) was significantly different than the Control condition (M=0.15, SD=0.98, p=0.001) and the Questionnaire condition (M=-0.14, SD=1.48, p=0.029). The Chat condition (M=-0.58, SD=1.26) also differed significantly from Control (p=0.001).

\textbf{\(\Delta\) Overwhelmed}: A Welch one-way ANOVA was conducted due to a violation of the homogeneity of variances assumption (Levene's test, p=0.047). There was no significant effect of intervention conditions on change in feeling overwhelmed at the p$<$.05 level [F(3,185.98)=2.24, p=0.085]. 

\textbf{\(\Delta\) Positive Emotion}: A Welch one-way ANOVA was conducted due to violation of the homogeneity of variances assumption (Levene's test, p=0.004). There was no significant effect of intervention conditions on change in positive emotion at the p$<$.05 level [F(3,183.84)=2.32, p=0.077].  

\textbf{\(\Delta\) Future Self-Continuity:} A Welch one-way ANOVA was conducted due to violation of the homogeneity of variances assumption (Levene's test, p$<$0.001). There was a significant effect of intervention conditions on the overall change in FSC at the p$<$.001 level [F(3,172.18)=8.57, p$<$0.001]. Post hoc comparisons using the Tukey HSD test indicated that the mean change for the Future You condition (M=0.42, SD=0.70) was significantly different than the Control condition (M=-0.00, SD=0.35, p=0.001), Chat condition (M=0.06, SD=0.42, p=0.001), and Questionnaire condition (M=0.22, SD=0.64, p=0.040). The Questionnaire condition also differed significantly from Control (p=0.040). Looking more closely at the specific sub-scales of future self-continuity:

\textbf{\(\Delta\) Future Self-Continuity - Similarity:} A Welch one-way ANOVA was conducted due to violation of the homogeneity of variances assumption (Levene's test, p$<$0.001). There was a significant effect of intervention conditions on change in FSC similarity at the p$<$.001 level [F(3,173.60)=7.45, p$<$0.001]. Post hoc comparisons using the Tukey HSD test indicated that the mean change for the Future You condition (M=0.58, SD=1.08) was significantly different than the Control condition (M=-0.06, SD=0.52, p=0.001), Chat condition (M=0.03, SD=0.56, p=0.001), and Questionnaire condition (M=0.13, SD=0.83, p=0.002).

\textbf{\(\Delta\) Future Self-Continuity - Vividness:} A Welch one-way ANOVA was conducted due to violation of the homogeneity of variances assumption (Levene's test, p$<$0.001). There was a significant effect of intervention conditions on change in FSC vividness at the p$<$.01 level [F(3,175.00)=4.38, p=0.005]. Post hoc comparisons using the Tukey HSD test indicated that the mean change for the Future You condition (M=0.47, SD=0.84) was significantly different than the Chat condition (M=0.15, SD=0.59, p=0.017) and Control condition (M=0.10, SD=0.49, p=0.006).

\textbf{\(\Delta\) Future Self-Continuity - Positivity: }A Welch one-way ANOVA was conducted due to violation of the homogeneity of variances assumption (Levene's test, p=0.005). There was a marginally significant effect of intervention conditions on change in FSC positivity at the p$<$.05 level [F(3,173.18)=2.55, p=0.057].

\textbf{\(\Delta\) Future Consideration:} A Welch one-way ANOVA was conducted due to a violation of the homogeneity of variances assumption (Levene's test, p=0.018). There was no significant effect of intervention conditions on change in composite future consideration at the p$<$.05 level [F(3,184.72)=1.19, p=0.314].

\textbf{\(\Delta\) Self-Esteem, \(\Delta\) Agency, and \(\Delta\) Optimism: }A Welch one-way ANOVA was conducted due to violation of the homogeneity of variances assumption (Levene's test, p=0.006). There was no significant effect of intervention conditions on change in composite self-esteem at the p$<$.05 level [F(3,183.36)=1.21, p=0.308]. For Agency, a Welch one-way ANOVA was conducted due to violation of the homogeneity of variances assumption (Levene's test, p=0.024). There was a significant effect of intervention conditions on change in composite agency at the p$<$.01 level [F(3,183.19)=4.68, p=0.004]. Post hoc comparisons using the Tukey HSD test indicated that the mean change for the Control condition (M=-0.12, SD=0.57) was significantly lower than the Chat condition (M=0.20, SD=0.67, p=0.009). For optimism, a Welch one-way ANOVA was conducted due to a violation of the homogeneity of variances assumption (Levene's test, p=0.006). There was no significant effect of intervention conditions on change in composite optimism at the p$<$.05 level [F(3,181.43)=1.50, p=0.217].

\textbf{\(\Delta\)  Self-Reflection and \(\Delta\) Insight:} A one-way ANOVA was conducted to compare the effect of intervention conditions on change in composite self-reflection. The assumption of homogeneity of variances was met (Levene's test, p=0.203). There was no significant effect of intervention conditions on change in composite self-reflection at the p$<$.05 level [F(3,376)=0.47, p=0.703].  For insight, a one-way ANOVA was conducted to compare the effect of intervention conditions on change in composite insight. The assumption of homogeneity of variances was met (Levene's test, p=0.783). There was no significant effect of intervention conditions on change in composite insight at the p$<$.05 level [F(3,376)=0.69, p=0.557].

\section{Discussion}
In this study, we developed an interaction-based intervention to explore how interacting with a realistic future self could benefit users. The study yielded two key findings: immediately following the intervention, participants who interacted with their AI-generated future self reported significantly (1) decreased negative affect, including lowered anxiety; and (2) increased future self-continuity. Future research should directly compare our interaction-based Future You intervention with validated, embodiment-based interventions. 

Overall, further work is needed to understand the specific psychological mechanisms related to future self-continuity. We showed that it is not enough for interactions to feel conversational to strengthen the positive affect users feel toward their future selves, indicating that the interaction may only be affecting user emotions in the present. Qualitatively, however, users reported that they liked their future self, expressing feelings of comfort, warmth, and solace when chatting with their AI future selves. Users also reported that the chat interaction was enjoyable. Most users indicated that although the chat interaction felt artificial, it did not feel insincere, and that the interaction was conversational. This intervention thus contributes to a growing recognition of the potential for positive emotional interactions between humans and AI-generated virtual characters.

Beyond just prompting users to reflect on their futures, our intervention demonstrated that an \textit{interaction} between users and their future selves allowed for significant improvements in reducing users' negative affect and future self-continuity. Users noted that even though they perceived the life of their future self as different from their present life, their values and beliefs seemed consistent. As such, our intervention may help to combat the end-of-history illusion (the tendency to underestimate how much one’s future self will change from their present selves) in helping users realize that although the details of their lives may change drastically, their core identities can remain true \cite{quoidbach2013end}. Recent work has established a link between future self-continuity and authenticity—the sense that one is in alignment with their true self—which in turn promotes feelings of meaning \cite{hong2024future}. Future work should seek to understand how the future self interaction might directly affect a user's sense of authenticity regarding their present and future selves and whether authenticity could serve as a moderator for the effectiveness of the intervention.

Moreover, envisioning the future may not be enough to persuade people to pursue their future goals. Our research revealed that future-oriented behavioral and cognitive variables such as agency, optimism, and future consideration were not significantly impacted by the intervention, nor were self-related outcome variables of self-esteem, self-reflection, and insight. Recent work has revealed that users are more likely to overcome the intention-behavior gap when they feel connected but dissimilar to their future selves \cite{ganschow2024feeling}. A more fine-grained intervention targeting specific aspects of future self-continuity should explore how each domain of future self-continuity may contribute differently to how people envision, feel, and enact their futures \cite{OYSERMAN202373}. Future work should couple our intervention with behavioral goals to isolate the specific autobiographical, dialogical, and visual features that give such interactive interventions an edge in fostering future self-continuity and future-oriented behaviors, including consideration of how interacting with a future self might help users construct coherent narrative identities, or reframe personal decisions as interpersonal negotiations between present and future selves \cite{barresi2002fromthe, adler2016incremental}. 

\section{Limitations \& Ethical Consideration}
Our work opens new possibilities for AI-powered, interactive future self interventions and is part of a larger conversation on the ethics of human-AI interaction and AI-generated media \cite{pataranutaporn2021ai, pataranutaporn2022ai, prasongpongchai2024interactive, danry2022ai, dunnell2024ai}. There are many potential misuses of AI-generated future selves to be mindful of, including but not limited to: inaccurate depiction of the future in a way that harmfully influences present behavior, endorsement of negative behaviors, and the potential of over-reliance on such systems to assist decision-making \cite{pataranutaporn2021ai}. We call for researchers to futher investigate these possibilities and ensure the ethical use of futures thinking technology such as Future You. 

\section{Conclusion}
Our Future You intervention demonstrates a real-time, web-based intervention that shows promise in helping users build a closer relationship with their ideal future self—in a single, brief session. By chatting with a relatable yet virtual version of their future selves via a personalized large language model and age-progressed portrait, users experienced reduced negative emotions such as anxiety and unmotivated feelings and significantly increased future self-continuity. These findings highlight the potential of an AI-generated future self to help motivate and support the envisioning of one's future. We hope that our work will inspire more research in human-computer interaction that focuses on fostering future self-continuity and healthier, long-term decision-making and wellbeing.

\bibliographystyle{IEEEtran}
\bibliography{Reference}

\end{document}